\documentclass{article}
\usepackage{amssymb,amsfonts,amsthm}
\usepackage{graphicx}
\usepackage{subfigure}
\usepackage{amssymb}
\usepackage{amsthm}
\usepackage{amsmath}
\usepackage{bm}             
\usepackage[mathscr]{eucal}

\usepackage{cancel}
\usepackage{outbraces}

\usepackage{psfrag}

\usepackage{wasysym}

\newcommand{\E}[2]{E_{#1}{}^{#2}}

\newcommand{\N}[2]{\mathcal{N}^{#1}{}_{#2}}

\newcommand{\cT}{\mathcal{T}}
\newcommand{\cS}{\mathcal{S}}
\newcommand{\cA}{\mathcal{A}}
\newcommand{\cB}{\mathcal{B}}
\newcommand{\cO}{\mathcal{O}}

\theoremstyle{plain}

\newtheorem*{conjecture}{Conjecture}

\theoremstyle{remark}

\def\pmb#1{\setbox0=\hbox{$#1$}%
  \kern-.025em\copy0\kern-\wd0
  \kern.05em\copy0\kern-\wd0
  \kern-.025em\raise.0433em\box0}
\def\pmbs#1{\setbox0=\hbox{$\scriptstyle #1$}%
  \kern-.0175em\copy0\kern-\wd0
  \kern.035em\copy0\kern-\wd0
  \kern-.0175em\raise.0303em\box0}

\def\parb{\pmb{\partial}}

\def\be{\begin{equation}}
\def\ee{\end{equation}}
\def\bea{\begin{eqnarray}}
\def\eea{\end{eqnarray}}
\def\lb{\label}

\def\gam{\gamma}
\def\d{\delta}
\def\eps{\epsilon}

\def\Sig{\Sigma}

\def\Udot{\dot{U}}

\def\cn{{\cal N}}

\def\bOm{\mbox{\boldmath $\Omega$}}
\def\bna{\mbox{\boldmath $\nabla$}}

\def\ptl{\partial}

\def\hsp5{\hspace{5mm}}
\newcommand{\sfrac}[2]{{\textstyle{#1\over#2}}}
\def\case#1/#2{\textstyle\frac{#1}{#2}}

\begin{document}


\title{\bf The Nature of Generic Cosmological Singularities\footnote{Invited talk at the 11th
Marcel Grossmann Meeting on Recent Developments in General
Relativity, Berlin, Germany, 23-29 July 2006.}}
\author{
Claes Uggla\thanks{Electronic address: {\tt Claes.Uggla@kau.se}}\\
\\
{\em Department of Physics, University of Karlstad,}\\
{\em S-651 88 Karlstad, Sweden}}

\date{\normalsize{}}
\maketitle

\begin{abstract}
The existence of a singularity by definition implies a preferred
scale---the affine parameter distance from/to the singularity of a
causal geodesic that is used to define it. However, this variable
scale is also captured by the expansion along the geodesic, and this
can be used to obtain a regularized state space picture by means of
a conformal transformation that factors out the expansion. This
leads to the conformal `Hubble-normalized' orthonormal frame
approach which allows one to translate methods and results
concerning spatially homogeneous models into the generic
inhomogeneous context, which in turn enables one to derive the
dynamical nature of generic cosmological singularities. Here we
describe this approach and outline the derivation of the
`cosmological billiard attractor,' which describes the generic
dynamical asymptotic behavior towards a generic spacelike
singularity. We also compare the `dynamical systems
picture' resulting from this approach
with other work on generic spacelike singularities: the
metric approach of Belinskii, Lifschitz, and Khalatnikov, and the
recent Iwasawa based Hamiltonian method used by Damour, Henneaux,
and Nicolai; in particular we show that the cosmological billiards
obtained by the latter and the cosmological billiard attractor form
complementary `dual' descriptions of the generic asymptotic dynamics
of generic spacelike singularities.
\end{abstract}

\newpage

\section{Introduction}\label{intro}
The singularity theorems tell us that singularities occur under very
general circumstances in general relativity (GR), but they say
little about their nature---detailed insights require further
analysis of Einstein's field equations. Of particular interest is
the structure of generic singularities; it is probably fair to say
that there have historically been three `dominant' approaches to
address this issue: (i) the metric approach used by Belinskii,
Lifshitz, and Khalatnikov~\cite{lk63,bkl70,bkl82} (henceforth these
authors, and their work, are referred to as BKL), (ii) the
Hamiltonian approach, and (iii) the dynamical systems approach, which
will be the main topic of the present paper, but let us begin with a
brief historical review.

During the sixties, seventies, and early eighties, BKL performed a
heuristic analysis that eventually resulted in the conjecture that a
generic singularity is {\em spacelike\/}, {\em
local\/},\footnote{Local means that near the singularity the spatial
points essentially `decouple,' in the sense that the dynamical
evolution is asymptotically governed by ordinary differential
equations with respect to time.} {\em vacuum dominated\/}, and {\em
oscillatory\/}, when the matter source is a perfect fluid with a
radiation equation of state, or more precisely: the time evolution
of a solution in the vicinity of a generic singularity is described
by a sequence of generalized Kasner solutions,\footnote{Contrary to
what the name implies, the generalized Kasner solution is not a
solution to Einstein's equations; instead it is the line element one
obtains if one replaces the constants in the vacuum Bianchi type I
solution with spatially dependent functions.} where one generalized
Kasner solution is linked to the next by a generalized vacuum
Bianchi type II solution, leading to an oscillatory behavior,
discretely described by a chaotic Kasner
map~\cite{khaetal85,bar82,chebar83}. To reach these conclusions BKL
(i) employed synchronous coordinates, i.e. Gaussian normal
coordinates, such that the singularity occurred simultaneously, (ii)
considered certain spatially homogeneous (SH) metrics, (iii)
replaced constants with spatial functions, (iv) inserted the
resulting expressions into Einstein's field equations as a starting
point for a perturbative expansion which yielded a consistency
check, which, after some difficulties, led to the above scenario.

In the late sixties and early seventies the BKL picture obtained
heuristic support from results obtained by the Hamiltonian approach
to asymptotic Bianchi type IX dynamics developed by Misner and
Chitr\'e~\cite{mis69,grav73,chi72}. In one variation of this
approach the dynamical evolution towards the singularity is
described in terms of free motion in an abstract flat Lorentzian
`minisuperspace'~\footnote{Superspace is the space of spatial
metrics; minisuperspace is superspace restricted to the SH case.}
surrounded by potential walls, or alternatively by means of a
spatial projection in minisuper space which leads to free motion
inside a potential well described by moving walls; for further use
and developments of this picture, see the review by
Jantzen~\cite{jan01} and~\cite{waiell97} [Chapter 10]. In another
version of the Hamiltonian approach Misner and Chitr\'e{
}~\cite{grav73,chi72} described the asymptotic dynamics by
projecting the motion in minisuperspace onto a region in hyperbolic
space bounded by sharp walls, forming a `billiard table',
see~\cite{grav73,chi72}. The Hamiltonian approach was later
generalized so that the general inhomogeneous case could also be
studied, culminating in the recent work by Damour, Henneaux, and
Nicolai (aimed primarily at a broader string theoretic context),
see~\cite{dametal03,damnic05} and references therein.

Despite the indisputable ingenuity and power of the BKL and the
Hamiltonian approaches, they have unfortunately yielded rather vague
statements from a rigorous mathematical point of view, and this has
led to considerable debate. But there exist exceptions in
non-oscillatory cases: rigorous results have been obtained by
Moncrief, Isenberg, Berger, and collaborators, who also obtained
numerical support for the general basic BKL picture,
see~\cite{beretal98}, and~\cite{ber02} for a review and additional
references; note also the successful use of so-called Fuchsian
methods as regards theorems about asymptotic non-oscillatory
dynamics, particularly the result by Andersson and
Rendall~\cite{andren01} for the general case of a stiff fluid or
massless scalar field.

The last decade has seen a remarkable theoretical development in SH
cosmology, largely in connection with the book ``Dynamical Systems
in Cosmology''~\cite{waiell97}. The dynamics of e.g.\ perfect fluid
models are now to a large extent understood in terms of rigorous
statements, of which many have been elevated to mathematical
theorems. Notably, Ringstr\"om obtained the first theorems about
oscillatory behavior for Bianchi type VIII and, more substantially,
type IX~\cite{rin00,rin01} models. The progress in this area is due
to the use of dynamical systems techniques applied to the `dynamical
systems approach' to SH cosmology, see~\cite{waiell97,col03}.

In an attempt to broaden the success to the general inhomogeneous
case, Uggla, van Elst, Wainwright, and Ellis~\cite{uggetal03}
introduced a dynamical systems formulation for Einstein's field
equations without any symmetries. This led to a description of
generic asymptotic dynamics towards a generic spacelike singularity
in terms of an attractor, which resulted in mathematically precise
conjectures. Furthermore, this formulation served as a basis for
numerical investigations of generic singularities, which yielded
additional support for the expected generic picture as well as the
discovery of new phenomena and subsequent
refinements~\cite{gar04,lim04,limetal04,andetal05,limetal06}.
Recently this work was given a more sound geometric
foundation~\cite{rohugg05}, and further steps were taken by Heinzle,
Uggla, and R\"ohr~\cite{heietal07} to sharpen and substantiate exact
rigorous mathematical statements about generic asymptotic dynamics
towards a generic spatial singularity---a development we focus on
below.

\section{The conformal Hubble-normalized dynamical systems approach}\label{confsec}
Let us consider a generic spacelike singularity, for simplicity
located in the past, which motivates referring to it as a
`cosmological singularity.' In this `cosmological' context it is
natural to focus on temporal aspects and investigate spacetime
changes along a timelike reference congruence that originates from
the singularity. Let us assume that we are sufficiently close to the
singularity so that the expansion $\theta=\bna_a\,u^a$ is positive,
where $u^a$ is the future-directed unit tangent vector field of the
reference congruence and where $\bna_a$ is the covariant derivative
associated with the physical metric ${\bf g}$.

Although the singularity theorems say little about the nature of
singularities, the very definition of a singularity and the one
dynamical input that goes into the theorems---the Raychaudhuri
equation for the expansion\footnote{In the case of timelike
geodesics; in the null geodesic case an analogous equation plays a
similar role.}---provide clues to how one might continue in the
quest to understand what happens in the vicinity of a generic
spacelike singularity. The existence of a singularity by definition
implies a prominent variable scale---the affine parameter distance
from/to the singularity of a causal inextendible geodesic that is
used to define it, or, alternatively, the expansion (cf. FRW
cosmology where the Hubble variable $H=\sfrac{1}{3}\,\theta$ is
frequently used to obtain a characteristic time scale, $H^{-1}$,
from the initial singularity). However, the key role played by the
Raychaudhuri equation in the singularity theorems suggests that the
expansion is particularly important---especially so in the case of a
generic spacelike singularity where the expansion is expected to
blow up, so that Einstein's field equations break down.

Causal properties constitute an important feature in
GR---particularly in the derivation of the singularity
theorems---and there are good reasons to believe that asymptotic
causal properties are absolutely crucial in the case of a generic
spacelike singularity: We expect a generic spacelike singularity to
be a scalar curvature singularity,\footnote{A non-scalar curvature
singularity requires fine tuning.} associated with increasing ultra
strong gravity that focuses light in all directions as the
singularity is approached, leading to {\em asymptotic silence\/},
which we define as the formation of particle horizons that shrink to
zero size in all directions along any time line that approaches the
singularity, thus increasingly prohibiting
communication;\footnote{One may also refer to this as an
anti-Newtonian limit, since, loosely speaking, the light cones are
flattened out onto spacelike surfaces in the Newtonian limit, while
they here `collapse' onto time lines.} for various special examples
of asymptotic silence and asymptotic silence-breaking, see Lim et
al.~\cite{limetal06}.

In relativity, due to the causal structure that links space and
time, there only exists a single dimensional unit, leaving us with
one dimensional scale. The above discussion suggests that one should
adapt to the asymptotic `dominant' scale and causal properties and
asymptotically {\em regularize\/} Einstein's equations by factoring
out the expansion or, equivalently, the Hubble variable
$H=\sfrac{1}{3}\,\theta$, so that one obtains dimensionless state
space variables that take {\em finite\/} values and capture the
`essential' dynamics: the geometric and natural way to do this is by
means of a conformal transformation; since ${\bf g}$ and $H^{-2}$
have dimensional weight [`length']$^2$ (or equivalently
[`time']$^2$), the appropriate transformation is given by
\begin{equation}
{\bf g} = H^{-2}\,{\bf G}\:,
\end{equation}
where ${\bf G}$ is a dimensionless unphysical metric.

To obtain a formulation that reduces to the Hubble-normalized
dynamical systems approach in SH cosmology, used in
e.g.~\cite{waiell97}, as a special case, we introduce: a {\em
conformal `Hubble-normalized' orthonormal frame\/},
\begin{equation} \label{defconfon}
{\bf g} = H^{-2}\,{\bf G} = H^{-2}\,\eta_{ab}\,\bOm^a\,\bOm^b\:,
\end{equation}
where $\eta_{ab}={\rm diag}(-1,1,1,1)$, and $a,b=0,1,2,3$; conformal
orthonormal vector fields $\parb_a$ dual to $\bOm^a$, i.e.,
$\langle\,\bOm^a,\,\parb_b\,\rangle = \delta^{a}{}_{b}$; a
non-rotating timelike reference congruence---asymptotically
conformally geodesic along a generic time line\footnote{A special
example of such a reference congruence is the synchronous time
choice used by BKL.}---to which we adapt the frame, i.e., $\parb_0=
H^{-1}{\bf u}$ is aligned tangentially to the time lines of the
reference congruence, and the shift vector is therefore set to zero.

This then yields
\begin{equation} \label{13cONfr}
\parb_{0} = \cn^{-1}\,\ptl_{x^0} =
\cn^{-1}\,\ptl_{0}\ , \qquad
\parb_{\alpha} = E_{\alpha}{}^{i}\ptl_{x^i} = E_{\alpha}{}^{i}\ptl_{i} \ ,
\qquad \alpha = 1,2,3 \ ; \quad i = 1,2,3\, ,
\end{equation}
where $\cn$ is the conformal lapse and $E_{\alpha}{}^{i}$ are the
conformal spatial frame vector components, related to corresponding
objects of ${\bf g}$ by
\begin{equation}\label{framedef}
\cn=H\,N\:,\qquad  E_{\alpha}{}^{i} = e_{\alpha}{}^{i}/H\:.
\end{equation}
Note that the components $E_\alpha{}^i$ are associated with the {\em
contravariant\/} spatial 3-metric of ${\bf G}$:
$G^{ij}=\delta^{\alpha\beta}E_{\alpha}{}^{i}E_{\beta}{}^{j}$; for
further details, see~\cite{rohugg05}.

The next ingredient in the `conformal Hubble-normalized orthonormal
frame approach' is the commutators of the dimensionless conformal
vector fields $\parb_a$:
\begin{align}
\label{ccomts0} [\,\parb_{0}, \parb_{\alpha}\,] & =
\dot{U}_{\alpha}\,\parb_{0} + (q\,\d_{\alpha}{}^{\beta} -
\Sig_{\alpha}{}^{\beta} -
\eps_{\alpha}{}^{\beta}{}_{\gam}\,R^{\gam})\,\parb_{\beta}\,,\\
\label{ccomt2} [\,\parb_{\alpha}, \parb_{\beta}\,] & =
(2A_{[\alpha}\,\d_{\beta]}{}^{\gam} +
\eps_{\alpha\beta\delta}\,N^{\delta\gam})\,\parb_{\gam}\, \:,
\end{align}
were $\Sig_{\alpha\beta}$, $\Udot_\alpha$, $R_\alpha$ are the shear,
acceleration, and Fermi rotation (which describes how the frame
rotates w.r.t. a Fermi propagated frame), respectively, all
associated with $\parb_0$ and ${\bf G}$; $N^{\alpha\beta}$,
$A_\alpha$ are spatial commutator functions that describe the
spatial three-curvature of ${\bf G}$ (see~\cite{waiell97,rohugg05}
where the analogous non-normalized objects are described); the
object $q$ is the deceleration parameter associated with the
physical spacetime and ${\bf u}$, but $q$ can also be interpreted
geometrically in terms of the expansion $\Theta$ of $\parb_0$
according to $q=-\frac{1}{3}\Theta$.

The only variable that carries dimension is $H$, and hence, for
dimensional reasons, the equations associated with $H$ decouple:
\be \label{Hubblenorm}
\parb_0 H = -(1+q)H\, ,\qquad \parb_\alpha H =
-r_\alpha\,H\, ,\ee
where the equations also serve to define $q$ and $r_\alpha$. The
dimensionless field equations yield a coupled system of evolution
equations and constraints for the variables
\be {\bf X}=(E_{\alpha}{}^{i})\oplus{\bf S}\oplus{\bf M}\:, \qquad
{\bf S}:=(\Sigma_{\alpha\beta},A_\alpha,N_{\alpha\beta})\:, \ee
where ${\bf X}$ is the state vector which describes the state space
and where ${\bf M}$ represents matter variables relevant for the
matter source one is interested in, however, from now on, except in
the concluding remarks, we will for simplicity restrict ourselves to
the vacuum case and hence the state space is described by the state
vector ${\bf X}=(E_{\alpha}{}^{i})\oplus{\bf S}$. In addition to
these variables there also exist gauge variables ${\bf X}_G$: the
$R_\alpha$ Fermi rotation variables are gauge variables associated
with the choice of frame; $\Udot_\alpha$ is regarded as a gauge
variable determined by the choice of ${\cal N}$ (or equivalently
$N$). The quantities $r_\alpha$ are generically determined by the
Codazzi constraint, however, for certain choices of ${\cal N}$ one
can derive an evolution equation for $r_\alpha$, and it may then be
advantageous to include $r_\alpha$ in ${\bf S}$,
see~\cite{uggetal03}; the deceleration parameter $q$ is determined
algebraically by the Raychaudhuri equation, but for certain time
choices it is possible and useful to also elevate $q$ to a dependent
variable, see~\cite{gar04}.

The conformally Hubble-normalized dimensionless system of coupled
partial differential equations are obtained from the above
Hubble-normalized commutator equations, which define ${\bf S}$ and
${\bf X}_G$ in terms of derivatives of the metric; the Jacobi
identities, which act as integrability conditions for going over to
an essentially first order formulation; and Einstein's field
equations~\cite{rohugg05}---the equations can be divided into gauge
equations, evolution equations, and constraint equations. The
evolution equations for the vacuum case can schematically be written
on the form
\be\label{evoleqs}
\parb_{0}E_{\alpha}{}^{i} =
F_{\alpha}{}^{\beta}\,E_{\beta}{}^{i}\,,\qquad
\parb_{0}{\bf S}={\bf P}\:, \ee
where $F_{\alpha}{}^{\beta}$ and ${\bf P}$ involve $\parb_\alpha,
{\bf S}, {\bf X}_G$ (in the matter case one adds matter equations
which yield evolution equations that schematically can be written as
$\parb_{0}{\bf M}={\bf Q}$, but apart from obvious modifications the
system takes the same form as in the vacuum case). Note that the
Hubble-normalized dimensionless equation system carries the
essential dynamical content, since once it is integrated the
solution can be inserted into the decoupled dimensional equations
for $H$, which subsequently can be integrated.

To obtain a determined system of evolution equations one needs to
specify the spatial frame. There exist many useful spatial frame
choices, e.g., the Fermi choice $R_\alpha=0$; the choice used
in~\cite{uggetal03}; the $SO(3)$ choice used by Benini and Montani
as starting point for a `Misner/Chitre' billiard
analysis~\cite{benmon04}; and the so-called Iwasawa choice used by
Damour, Henneaux, and Nicolai~\cite{dametal03}. To establish contact
with this latter work we will from now on use the Iwasawa choice,
however, it is probably safe to say that no frame choice will cover
all the features one may be interested in---each choice has its
advantages and disadvantages.

There exists a unique oriented orthonormal spatial frame
$\{\omega^{\alpha}\,|\,{\alpha}=1\ldots 3\}$, $\omega^{\alpha} =
e^{\alpha}_{\:\,i} dx^i$, such that $e^{\alpha}_{\:\,i} =
\sum_\beta\mathcal{D}^\alpha_{\:\,\beta}\,
\N{\beta}{i}$,\footnote{We here follow the summation conventions
in~\cite{dametal03}: summation of pairs of coordinate indices $i,
j,...$ is understood, whereas sums over the frame indices $\alpha,
\beta,...$ are written out explicitly when needed to avoid
confusion.} where $\mathcal{D}$ is a diagonal matrix, $\mathcal{D} =
\mathrm{diag}\big[\exp(-b^1),\exp(-b^2),\exp(-b^3)\big]$ (to avoid
confusion with the Greek frame indices $\alpha,\beta,\ldots$, we use
$b$ as the kernel letter for the diagonal degrees of freedom instead
of $\beta$, which was used in~\cite{dametal03}), and
$\N{{\alpha}}{i}$ a unit upper triangular matrix,
\begin{equation}
\Big(\,\N{{\alpha}}{i}\,\Big) =
\begin{pmatrix}
1 & \N{1}{2} & \N{1}{3} \\
0 & 1 & \N{2}{3} \\
0 & 0 & 1
\end{pmatrix} =
\begin{pmatrix}
1 & n_1 & n_2 \\
0 & 1 & n_3 \\
0 & 0 & 1
\end{pmatrix} \:;
\end{equation}
$\N{\alpha}{i}$ can also be viewed as representing the Gram-Schmidt
orthogonalization of the spatial coordinate coframe $\{d x^i\}$.
This choice of frame leads to the so-called Iwasawa decomposition of
the spatial metric $g_{i j}$:
\begin{equation}
g_{i j} = \sum_{\alpha} \exp(-2 b^{\alpha}) \N{{\alpha}}{i}
\N{{\alpha}}{j}\:.
\end{equation}
Using an Iwasawa based conformal Hubble-normalized frame, which we
from now on do, leads to $\Sigma_{23}= -R_1,\Sigma_{31} =
R_2,\Sigma_{12} = -R_3, N_{33}=0$~\cite{heietal07}. This makes it
natural to introduce the notation
$\Sigma_{\alpha}=\Sigma_{\alpha\alpha}$, while we replace the
off-diagonal components of $\Sigma_{\alpha\beta}$ by $R_\alpha$
according to the above equation, i.e.,
\begin{equation}\lb{sigmamatrix}
\begin{pmatrix}
\Sigma_{11} & \Sigma_{12} & \Sigma_{13} \\
\Sigma_{21} & \Sigma_{22} & \Sigma_{23} \\
\Sigma_{31} & \Sigma_{32} & \Sigma_{33}
\end{pmatrix} =
\begin{pmatrix}
\Sigma_{1} & -R_3 & R_2 \\
-R_3 & \Sigma_{2} & -R_1 \\
R_2 & -R_1 & \Sigma_{3}
\end{pmatrix} \:.
\end{equation}

\section{The silent boundary}\label{silbound}
In Section 4 in~\cite{uggetal03} it is shown that, generically,
asymptotic silence is connected with the property that
\[
\E{\alpha}{i} \rightarrow 0\:,
\]
while we define the dynamical evolution along a time line to be
\textit{asymptotically local} if
\begin{equation}
\label{asylocdef} E_{\alpha}{}^{i}\rightarrow 0 \:, \quad
\parb_{\alpha} (\bm{S}, r_\beta, \Udot_\beta) \rightarrow 0\:, \quad
(r_\alpha, \Udot_\alpha) \rightarrow 0 \:.
\end{equation}
towards the singularity. Asymptotic silence and asymptotic local
dynamics are closely related concepts, but not necessarily the same.
Numerical studies suggest that the dynamical evolution along generic
time lines towards generic asymptotically silent singularities is
asymptotically local, which corresponds to inhomogeneities being
shifted outside shrinking horizons faster than they grow
($E_{\alpha}{}^{i}$ goes to zero faster than $\partial_i (\bm{S},
\log(H),\log({\cal N}))$ may grow), however, numerics also suggest
that there may be {\em special\/} time lines for which this is not
true, which motivates the above
distinction~\cite{gar04,andetal05,heietal07}. Below, apart from in
the concluding remarks, we will discuss the generic case where the
evolution is assumed to be asymptotically local.

Consider Einstein's field equations in the conformal
Hubble-normalized approach, see Eq.~\eqref{evoleqs}, then it follows
that
\begin{equation}\label{silbou}
\E{\alpha}{i} = 0
\end{equation}
defines an invariant subset on the boundary of the state space,
characterized by the state vector $\bm{S}$; we refer to this
invariant subset as the \textit{silent boundary}. Since
$\parb_\alpha = \E{\alpha}{i}\partial_i$, Eq.~\eqref{silbou}
implies that the equations \textit{on} the silent boundary reduce to
a system of ordinary differential equations; \textit{the silent
boundary can therefore be visualized as an infinite set of copies,
parametrized by the spatial coordinates, of a finite dimensional
state space}, with $\bm{S}$ as state vector.

The equations on the silent boundary are identical to the equations
for $\bm{S}$ for spatially self-similar or SH
models~\cite{andetal04}, where the SH case is obtained by setting
\begin{equation}\label{SHsilentdef}
\E{\alpha}{i} =0 \quad\text{ and } \quad \Udot_\alpha = 0\,,\:
r_\alpha = 0\:.
\end{equation}
The particular importance of the `SH silent boundary' in the study
of generic spacelike singularities stems from its connection with
asymptotically local dynamics; compare Eqs.~\eqref{asylocdef}
and~\eqref{SHsilentdef}. Moreover, since the field equations in the
conformal Hubble-normalized orthonormal frame approach are regular
towards the singularity, we can \textit{extend} the state space to
include the silent boundary in our analysis of the dynamical system.
This is highly advantageous, since our above reasoning leads to the
conjecture that solutions in the full state space generically
asymptotically approach and \textit{shadow} the solutions on the SH
silent boundary.

The SH silent boundary consists of a number of invariant subsets
(`components') that are connected by parts of their boundary only,
where the precise structure depends on the choice of frame.
Utilizing an Iwasawa frame leads to equations on each
component which are the same as those for Bianchi types I--VII, but there
is no component associated with Bianchi type VIII or IX. The reason
for this is that the Iwasawa frame is incompatible with the symmetry
adapted frames of these models.\footnote{If one uses an Iwasawa
frame, Bianchi type VIII and IX solutions appear as inhomogeneous
solutions---with their symmetries hidden---in the full interior
state space $\bm{X}$.} The most general solutions on the SH silent
`Iwasawa' boundary---and these models are as general as the more
well known Bianchi type VIII and IX models---are the general Bianchi
type VI$_{-1/9}$ solutions~\cite{hewetal03}, and it is thus these
models that are of interest as the simplest exact SH example
exhibiting generic features in an Iwasawa context.

The asymptotic ODE structure induced by asymptotic local dynamics
makes it possible to \textit{re-parameterize individual time lines},
for which the spatial coordinates $x^i$ are fixed, so that
\be
\parb_{0}f = -df/d\tau
\ee
holds along a given time line, which allows us to study the
asymptotic dynamics along time lines by means of finite dimensional
dynamical systems techniques.\footnote{This is analogous to the
dynamical systems treatment of the so-called silent models,
see~\cite{matetal94,bruetal95} and~\cite{waiell97} [Chapter 13]; in
the present context it is irrelevant that the only nontrivial
spatially inhomogeneous non-rotating silent solutions without a
cosmological constant turned out to be the Szekeres dust models,
see~\cite{hveetal97,wylber06}, and references therein.} In the above
equation $f(x^0,x^i)$ is any variable occurring in an ODE;
$\tau(x^0,x^{i}):= -\log(\ell/\hat{\ell})$ is a `local time
function' directed towards the singularity; $\ell$ is related to the
the determinant $g$ of the physical spatial metric according to
$\ell=g^{1/6}$; $\hat{\ell}$ is an initial value that may vary for
different spatial points, i.e., $\hat{\ell}$ is a spatially
dependent function, and throughout we will use the convention that
hatted objects refer to quantities that are functions of the spatial
coordinates alone. To obtain the solution in the chosen time
coordinate $x^0$, one integrates the relation $dx^0 =
-\cn^{-1}d\tau$ so that
\begin{equation} \label{ttaurel} x^0 = {\hat x}^0(x^{i}) -
\int_{\tau_0}^{\tau}\cn^{-1}({\tau}',x^{i})d{\tau}'\, .
\end{equation}

\section{The billiard attractor}\label{attractor}
If a generic solution approaches the SH part of the silent boundary,
then due to the regularity of the equations, it will asymptotically
approach the attractor on this subset. Our experience with SH
cosmology suggests that this attractor is somehow connected with the
Bianchi type I and II subsets on the silent boundary, but that it
also depends on the choice of frame.

Dynamical systems investigations usually begin by looking for fixed
points (equilibrium points) and then one subsequently linearizes the
equations at these points to obtain a local picture of dynamical
features. In the Iwasawa gauge there exists a representation of the
(vacuum) Bianchi type I, i.e. Kasner, solutions on the silent
boundary as a one-parameter set of fixed points, which play a key
role for the asymptotic dynamics; this set of fixed points is
referred to as the \textit{Kasner circle} $\mathrm{K}^{\ocircle}$,
and is determined by
\begin{equation}\label{Kasnercircledef1}
1-\Sigma^2 = A_{\alpha} = N_{\alpha\beta}= R_\alpha = 0 \:,
\end{equation}
where
$\Sigma^2=\frac{1}{6}\Sigma_{\alpha\beta}\Sigma^{\alpha\beta}$. It
follows from Eqs.~\eqref{sigmamatrix} and~\eqref{Kasnercircledef1} that
$\Sigma_{\alpha\beta}=
\mathrm{diag}[\hat{\Sigma}_{1},\hat{\Sigma}_{2},\hat{\Sigma}_{3}]$,
however, one usually represents the Kasner circle by the generalized
Kasner exponents $p_\alpha$:
\begin{equation}\label{Kasnercircledef2}
\Sigma_{\alpha\beta}=\text{diag}[{\hat \Sig}_{1},{\hat
\Sig}_{2},{\hat \Sig}_{3}]= \text{diag}[3p_1-1,3p_2-1,3p_3-1]\:,
\end{equation}
where we omit the hats on top of $p_\alpha$ in order to agree with
standard notation. Since $\mathrm{tr}\: \Sigma_{\alpha\beta} = 0$
and $\Sigma^2 =1$, the Kasner exponents satisfy the Kasner relations
\begin{equation}
p_1+p_2+p_3=1\:, \qquad p_1^2 + p_2^2 + p_3^2 = 1\:,
\end{equation}
which imply that $-\frac{1}{3} \leq p_\alpha \leq 1$ (or
equivalently, $-2\leq \Sigma_\alpha \leq 2$).

\begin{figure}[h]
\psfrag{a}[cc][cc]{$\Sigma_1$} \psfrag{b}[cc][cc]{$\Sigma_3$}
\psfrag{c}[cc][cc]{$\Sigma_2$} \psfrag{d}[cc][cc]{$T_1$}
\psfrag{e}[cc][cc]{$Q_2$} \psfrag{f}[cc][cc]{$T_3$}
\psfrag{g}[cc][cc]{$Q_1$} \psfrag{h}[cc][cc]{$T_2$}
\psfrag{i}[cc][cc]{$Q_3$} \psfrag{j}[cc][cc]{0}
\psfrag{k}[cc][cc]{$\sfrac{\pi}{3}$}\psfrag{l}[cc][cc]{$\sfrac{2\pi}{3}$}
\psfrag{m}[cc][cc]{$\pi$} \psfrag{n}[cc][cc]{$\sfrac{4\pi}{3}$}
\psfrag{o}[cc][cc]{$\sfrac{5\pi}{3}$}\psfrag{p}[cc][cc]{$2\pi$}
\psfrag{q}[cc][cc]{$u$} \psfrag{r}[cc][cc]{$\alpha$}
\psfrag{A}[cc][cc]{$N_2,R_3$} \psfrag{B}[cc][cc]{$N_1$}
\psfrag{C}[cc][cc]{$N_1,R_1$} \psfrag{D}[cc][cc]{$R_1,R_2$}
\psfrag{E}[cc][cc]{$R_1,R_2,R_3$}\psfrag{F}[cc][cc]{$N_2,R_2,R_3$}
 \centering{
       \includegraphics[width=0.45\textwidth]{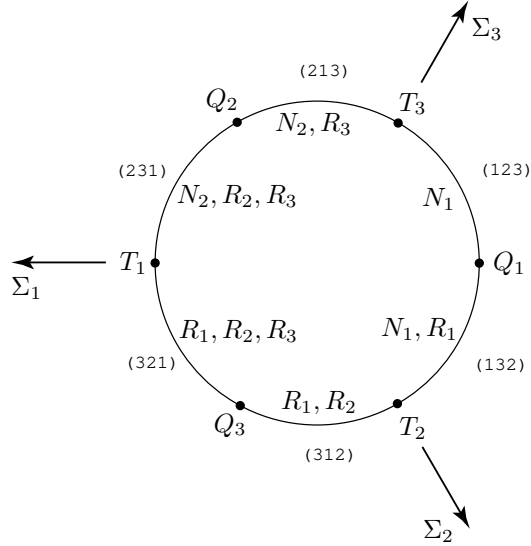} \\
        \caption{The Kasner circle of fixed points---sectors and unstable
        variables.}}
\label{Ktrig}
\end{figure}

The Kasner circle $\mathrm{K}^{\ocircle}$ can be divided into six
equivalent sectors, identifiable by means of permutations of the
spatial axes. Each sector can be characterized by an ordered
sequence of Kasner exponents; hence e.g. sector $(312)$ is defined
as the part of $\mathrm{K}^{\ocircle}$ where $p_3<p_1<p_2$, see
Fig.~\ref{Ktrig}. There are six special points on
$\mathrm{K}^{\ocircle}$ that are associated with solutions with
additional symmetries: the points $Q_\alpha$ correspond to the three
equivalent non-flat plane symmetric solutions, while the points
$T_\alpha$ correspond to the Taub representation of Minkowski
spacetime, see Fig.~\ref{Ktrig}.\footnote{For $Q_1$ we have
$(\hat{\Sigma}_1,\hat{\Sigma}_2,\hat{\Sigma}_3) = (-2,1,1)$ or
$(p_1,p_2,p_3) = (-\frac{1}{3},\frac{2}{3},\frac{2}{3})$; $T_1$ is
characterized by $(\hat{\Sigma}_1,\hat{\Sigma}_2,\hat{\Sigma}_3) =
(2,-1,-1)$ or $(p_1,p_2,p_3) = (1,0,0)$; $Q_2,Q_ 3,T_2,T_3$ are
obtained by cyclic permutations).}

Linearization of the dimensionless field equations at
$\mathrm{K}^{\ocircle}$ yields a set of ODEs that tells us that
$E_\alpha{}^i$, $N_{\alpha\beta}$ ($\alpha \neq \beta$) and
$A_\alpha$ belong to the stable subspace of each fixed point of
$\mathrm{K}^{\ocircle}$ (except at the Taub points; e.g.
$\partial_\tau E_\alpha{}^i = -3 (1 - p_\alpha) E_\alpha{}^i $),
while $R_\alpha$ and $N_1=N_{11}, N_2=N_{22}$ are stable or unstable
depending on the sector of $\mathrm{K}^{\ocircle}$ the point
$(p_1,p_2,p_3)$ lies in. Finally, the variables $\Sigma_\alpha =
\Sigma_{\alpha\alpha}$ belong to the center subspace, i.e., they are
constant to first order. The analysis of the stability of the Kasner
circle $\mathrm{K}^{\ocircle}$ is summarized in Fig.~\ref{Ktrig},
where the unstable variables are given for each sector of
$\mathrm{K}^{\ocircle}$.

The linear analysis suggests that the stable variables
$E_\alpha{}^i$ and $\bm{S}_{\mathrm{stable}}=(N_{\alpha\beta},
A_{\gamma})$ ($\alpha \neq \beta$) decay rapidly, and since a
generic solution is expected to spend an increasing amount of time
in a small neighborhood of the `non-flat' part of
$\mathrm{K}^{\ocircle}$, we are led to the conjecture that a generic
solution must approach the invariant \textit{oscillatory subset}
$\mathcal{O}$, determined by $\bm{S}_{\mathrm{stable}} = 0$ on the
SH silent boundary. This motivates decomposing the state
vector $\bm{S} = (\Sigma_{\alpha\beta}, A_{\alpha},
N_{\alpha\beta})$ of the SH silent boundary into a `stable' and an
`oscillatory' part:
\begin{equation}
\bm{S} = \bm{S}_{\mathrm{stable}} \oplus \bm{S}_{\mathrm{osc}}\qquad
{\rm where}\qquad \bm{S}_{\mathrm{osc}} = (\Sigma_\alpha,
R_{\alpha}, N_1, N_2)\:.
\end{equation}
The oscillatory subset $\mathcal{O}$, characterized by the state
vector $\bm{S}_{\mathrm{osc}}$, consists of components describing
various representations of Bianchi type I, II, $\mathrm{VI}_0$,
$\mathrm{VII}_0$ solutions. All the orbits, i.e. solutions, on these
components are heteroclinic orbits, i.e. they originate and end at
fixed points.

Since $\E{\alpha}{i}\rightarrow 0, \bm{S}_{\mathrm{stable}}
\rightarrow 0$, it is the asymptotic behavior of the remaining
oscillatory variables $\bm{S}_{\mathrm{osc}}$ that represents the
nontrivial asymptotic dynamics of a generic solution
$\bm{X}(\tau)$, at a generic spatial point $x^i$, as
$\tau\rightarrow \infty$. Moreover, due to the heteroclinic orbit
structure on $\mathcal{O}$, we expect $\bm{S}_{\mathrm{osc}}(\tau)$
to be increasingly accurately described by a {\em partition based on
a sequence of segments, where each segment is associated with a
heteroclinic orbit on $\mathcal{O}$ and where two subsequent
segments of $\bm{S}_{\mathrm{osc}}(\tau)$ are joined at a point of
$\mathrm{K}^{\ocircle}$\/}, yielding an oscillatory behavior. We
refer to this description of the orbit $\bm{S}_{\mathrm{osc}}(\tau)$
in the asymptotic regime as an \textit{asymptotic sequence of
$\mathcal{O}$-orbits} $\cA\cS_\cO$, or for brevity, as an
\textit{asymptotic sequence} (for an example of a part of an
asymptotic sequence, see Fig.~3($b$)). Note that the concept of
asymptotic sequences $\cA\cS_\cO$ generalizes, connects, and gives
precise meaning to BKL's `piecewise approximations'.

The heteroclinic orbits on $\mathcal{O}$ form a `heteroclinic orbit
puzzle' that governs asymptotically local dynamics, but an analysis
shows that it is only those orbits that connect points on
$\mathrm{K}^{\ocircle}$ that are of relevance for asymptotic generic
dynamics~\cite{heietal07}; these heteroclinic orbits can be regarded
as representing \textit{transitions} between different points on
$\mathrm{K}^{\ocircle}$.

\begin{figure}[ht]
$$
\begin{array}{ccc}
        \includegraphics[height=0.25\textwidth]{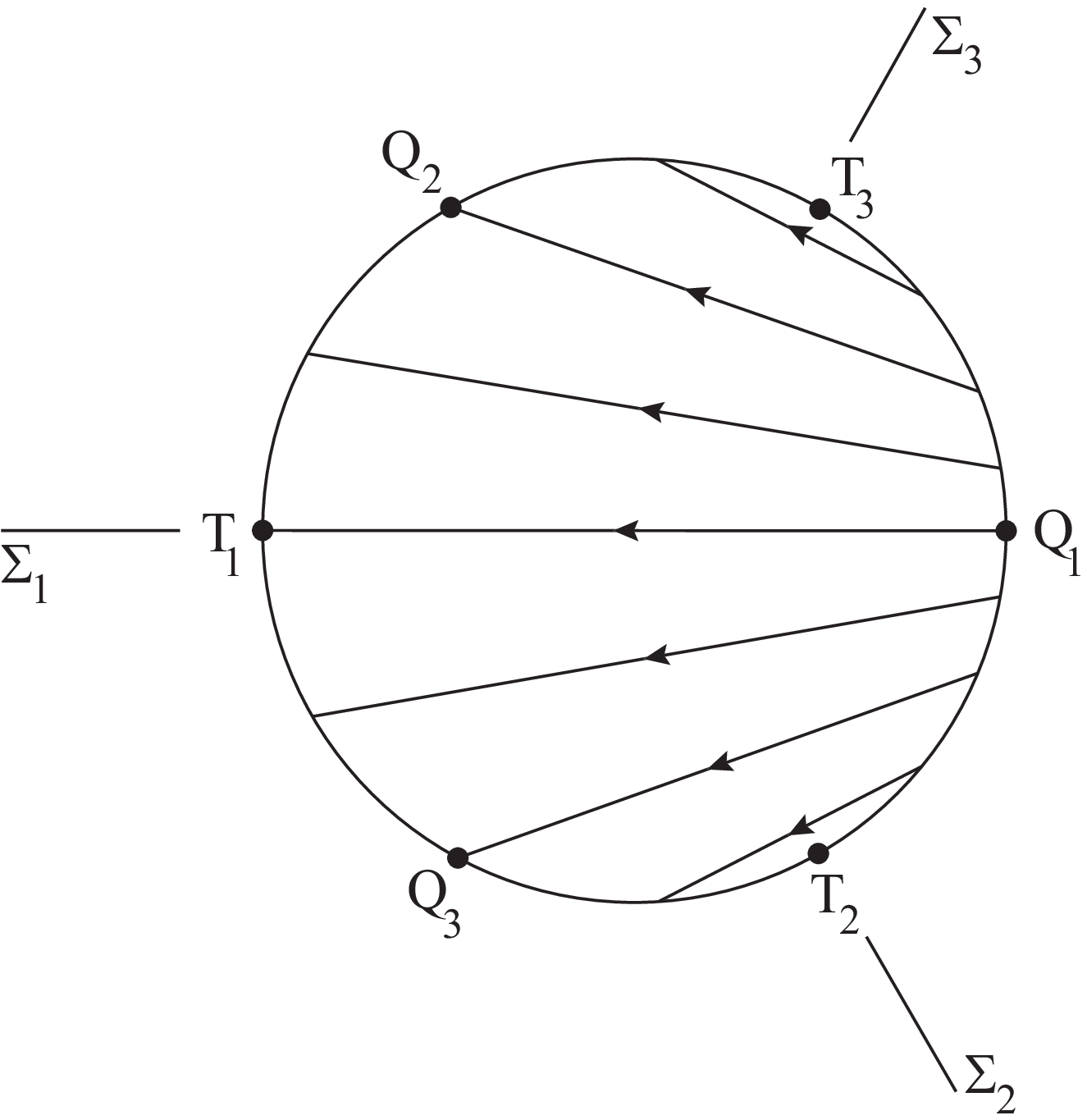}\qquad\qquad
&
        \includegraphics[height=0.25\textwidth]{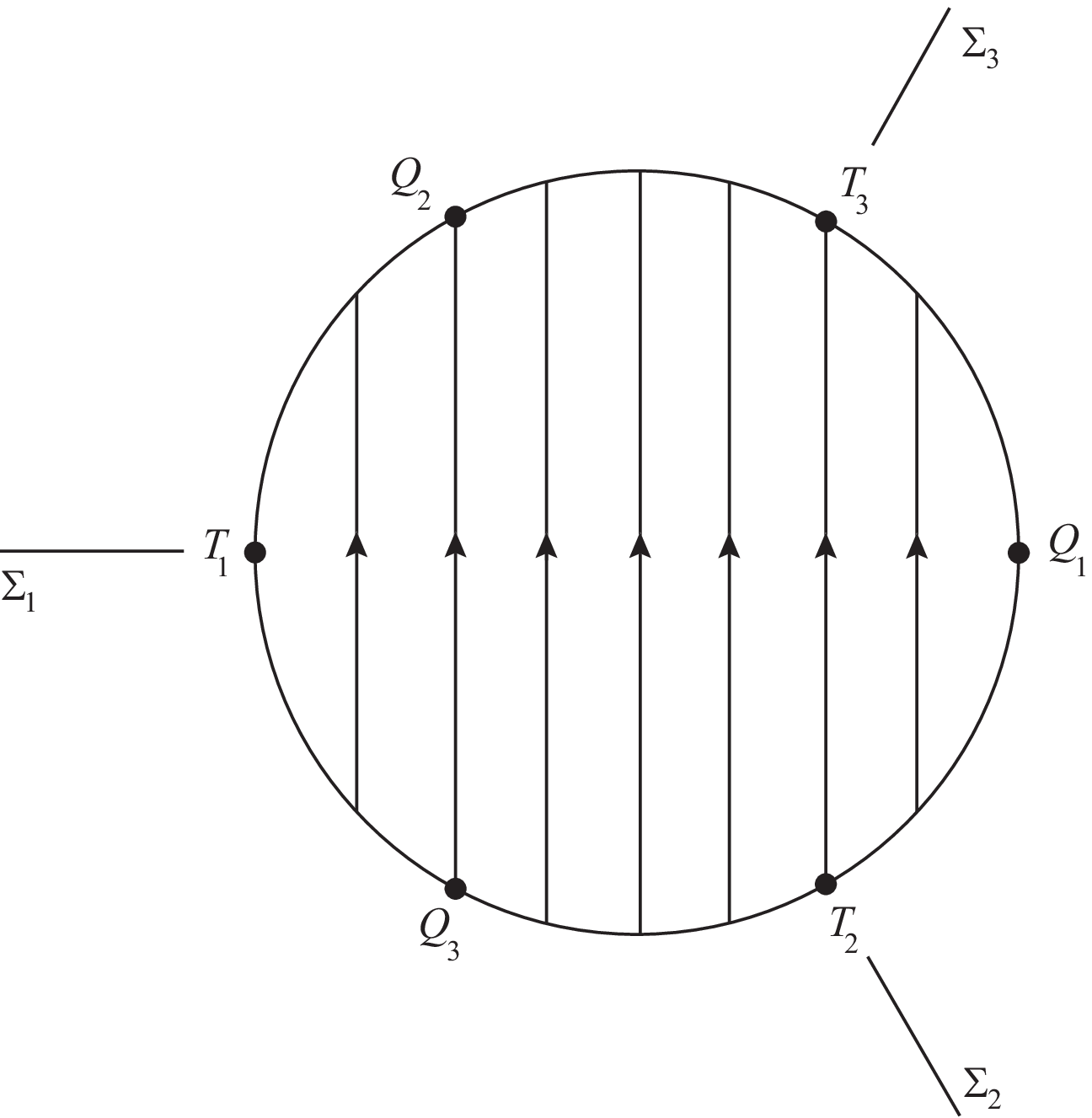}\qquad\qquad
&
        \includegraphics[height=0.25\textwidth]{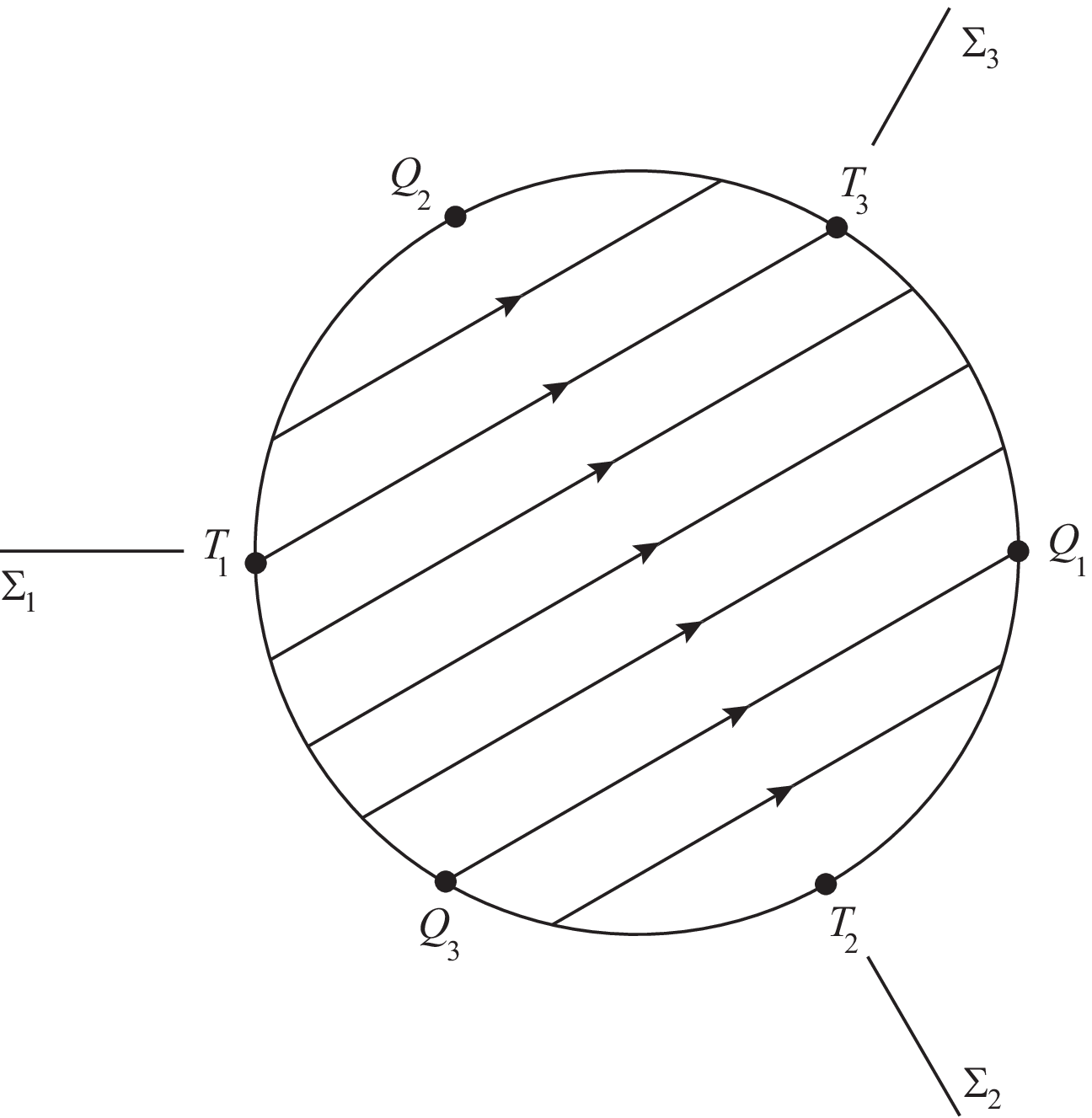}\\
\mathcal{T}_{N_1} & \mathcal{T}_{R_1} & \mathcal{T}_{R_3}
\end{array}
$$
\caption{Projections onto diagonal $\Sigma$-space of the single transitions
        $\mathcal{T}_{N_1}$, $\mathcal{T}_{R_1}$, $\mathcal{T}_{R_3}$
        on the billiard attractor for the Iwasawa frame.}
\label{singletrans}
\end{figure}

Further analysis shows that generic dynamics is associated with a
subset of $\mathcal{O}$ ---the \textit{billiard attractor} subset
$\mathcal{O}_{\mathcal{BA}}$~\cite{heietal07}, defined as
\begin{equation}\label{billiard}
\mathcal{O}_{\mathcal{BA}} =
\mathrm{K}^{\ocircle}\cup\cB_{N_1}\cup\cB_{R_1}\cup\cB_{R_3}\:,
\end{equation}
where $\cB_{N_1},\cB_{R_1},\cB_{R_3}$ are subsets on $\mathcal{O}$
characterized by $N_2=R_1=R_2=R_3=0$ (the silent Bianchi type II
subset associated with $N_1$ and a Fermi frame), $N_1=N_2=R_2=R_3=0$
(the silent Kasner subset in a frame that rotates w.r.t.\ a Fermi
propagated frame in the 2-3-plane), $N_1=N_2=R_1=R_2=0$ (the silent
Kasner subset in a frame that rotates w.r.t.\ a Fermi propagated
frame in the 1-2-plane), respectively; transitions, i.e.
heteroclinic orbits that connect points on $\mathrm{K}^{\ocircle}$,
associated with these subsets, denoted by $\mathcal{T}_{N_1}$
(called single curvature transitions), $\mathcal{T}_{R_1}$,
$\mathcal{T}_{R_3}$ (called single frame transitions), are depicted
as projections onto $\Sigma_\alpha$-space in Fig.~\ref{singletrans}.

An \textit{attractor sequence} $\cA_\cT$ is a sequence of
transitions on the billiard attractor $\mathcal{O}_{\mathcal{BA}}$,
i.e an infinite concatenation of single transitions $\cT_{N_1}$,
$\cT_{R_1}$, and $\cT_{R_3}$; to get an intuitive picture of
attractor sequences we refer to Fig.~\ref{singletrans} and give an
example of part of an attractor sequence in Fig. 3$(b)$.

The above leads to the formulation of the \textit{dynamical systems
billiard conjecture}~\cite{heietal07}:
\begin{conjecture}
The asymptotic dynamics of a generic time line of a solution of
Einstein's vacuum equations (expressed in an Iwasawa frame) that
exhibits a generic spacelike singularity is characterized as
follows:
\begin{itemize}
\item[\rm{(i)}] It is asymptotically silent and local.
\item[\rm{(ii)}] In the asymptotic limit the essential dynamics
is represented by an attractor sequence $\cA_\cT$ on the billiard
attractor $\mathcal{O}_{\mathcal{BA}}$.
\end{itemize}
\end{conjecture}

Let us comment on this, and in particular, on some of the features
and ingredients in the \textit{derivation} in~\cite{heietal07} that
led to the above conjecture:
\begin{itemize}
\item[\rm{(i)}] The suppression of a priori possible generic behavior
can be divided into two kinds: a) `dynamical' suppression, b)
`stochastical' suppression.
\item[\rm{(ii)}] The derivation of `suppression estimates' is based
on utilizing the fact that the asymptotic dynamical evolution of a generic
spatial point $x^i$ of a generic solution $\bm{X}(\tau)$ shadows a
sequence of orbits/transitions on $\mathcal{O}$ with an increasing
degree of accuracy. This leads to the concept of asymptotic
sequences, which yield a simple asymptotic dynamical description in
the context of the full extended state space picture.
\item[\rm{(iii)}] Although an orbit is increasingly accurately
approximated by a sequence of heteroclinic orbit segments,
decreasing errors nevertheless occur, and from a stochastic point of
view this is essential: they lead to a `randomization of
heteroclinic orbits,' and this is what makes a stochastic analysis
admissible.
\item[\rm{(iv)}] The stochastic analysis of sequences of transitions
relies on a generalization of BKL's concept of so-called eras;
sequences are partitioned into {\em small curvature phases\/}
(series of heteroclinic orbits close to the Taub points where the
curvature is small) and {\em large curvature phases\/} (series of
heteroclinic orbits far from the Taub points where the curvature is
large), where, stochastically, small curvature phases turn out to
dominate over large curvature phases.
\item[\rm{(v)}] The dynamical and stochastical analysis
leads to an estimation of decay rates which shows that, in the
context of a hierarchy of subsets, the asymptotic dynamical
evolution is restricted to subsets of subsets, to boundaries of
boundaries, descending from the full state space via the SH silent
boundary and the oscillatory subset down to the billiard attractor.
\item[\rm{(vi)}] Stochastic considerations are crucial in the derivation of
the billiard attractor. The possibility therefore exists that there are
solutions with different asymptotic behavior, but such solutions are
expected to form a set of measure zero in the space of all
solutions.
\item[\rm{(vii)}] The attractor is a local attractor in the full state
space---there exist open sets of solutions not approaching it (e.g.,
an open set of solutions that are almost flat everywhere).
\end{itemize}

The derivation of the billiard conjecture in~\cite{heietal07} is not
mathematically rigorous, but depends on arguments that are
heuristic---despite their being convincing. However, it is
reasonable to expect that if one attempts to obtain proofs, then
several concepts and methods introduced in~\cite{heietal07} will
play a prominent role. Let us give an example: The underlying
structures that allowed one to obtain mathematical proofs about the
attractor of the diagonal Bianchi type IX models in a Fermi
propagated frame~\cite{rin00,rin01} are specific for these models
and are not available in other cases; since it is such cases that
are relevant for the present general scenario, the diagonal Bianchi
type IX models are misleading. The analysis in~\cite{heietal07}
suggests that one has to know the (asymptotic) history of a solution
to unravel its asymptotic features; this causes a dilemma since this
requires that one finds the solution, which seems unlikely. However,
randomization makes it possible to stochastically examine the
cumulative effects of small and large curvature phases and this
allows one to estimate decay rates and give a description of what is
going to happen generically---statistical analysis is one example of
an ingredient that is likely to play a role in future proofs.

\section{`Duality' of Hamiltonian and dynamical systems billiards}\label{dual}
In the Hamiltonian approach by Damour, Henneaux, and
Nicolai~\cite{dametal03,damnic05} it is shown that the essential
generic asymptotic dynamics can be described by a Hamiltonian
cosmological billiard. In this picture the evolution of a spatial
point of a (generalized) Fermi propagated Kasner solution appears as
a geodesic in hyperbolic space, but this space is bounded by sharp
walls, and this leads to bounces, see Fig.~3($a$); the walls and
bounces are of two kinds: (i) frame/centrifugal/symmetry walls and
associated bounces; (ii) curvature walls and associated bounces.
Bounces of the former type merely result in axis permutations of a
Kasner solution, while bounces of the latter type lead to a change
in the Kasner state generated by a Bianchi type II solution.

In the dynamical systems formulation free motion in the Hamiltonian
picture corresponds to the points on the Kasner circle
$\mathrm{K}^{\ocircle}$; centrifugal bounces in the Hamiltonian
approach translate to single frame transitions $\cT_{R_1}$ and
$\cT_{R_3}$, while curvature bounces are connected with single
curvature transitions $\cT_{N_1}$. Here we observe `bounces' at the
fixed points on the Kasner circle $\mathrm{K}^{\ocircle}$, which act
as a `wall', and between bounces we have motion along straight lines
in (Euclidean) $\Sigma_\alpha$-space (transitions), see Fig.~3($b$).

\begin{figure}[ht]  \label{cosmobilliard}
\psfrag{a}[cc][cc]{$\gamma^1$}
\psfrag{b}[cc][cc]{$\gamma^3$}\psfrag{c}[cc][cc]{$\gamma^2$}
\psfrag{j}[cc][cc]{$\Sigma_1$} \psfrag{k}[cc][cc]{$\Sigma_3$}
\psfrag{l}[cc][cc]{$\Sigma_2$} \psfrag{d}[cc][cc]{$T_1$}
\psfrag{e}[cc][cc]{$Q_2$} \psfrag{f}[cc][cc]{$T_3$}
\psfrag{g}[cc][cc]{$Q_1$} \psfrag{h}[cc][cc]{$T_2$}
\psfrag{i}[cc][cc]{$Q_3$}
$$
\begin{array}{cc}
\includegraphics[width=0.35\textwidth]{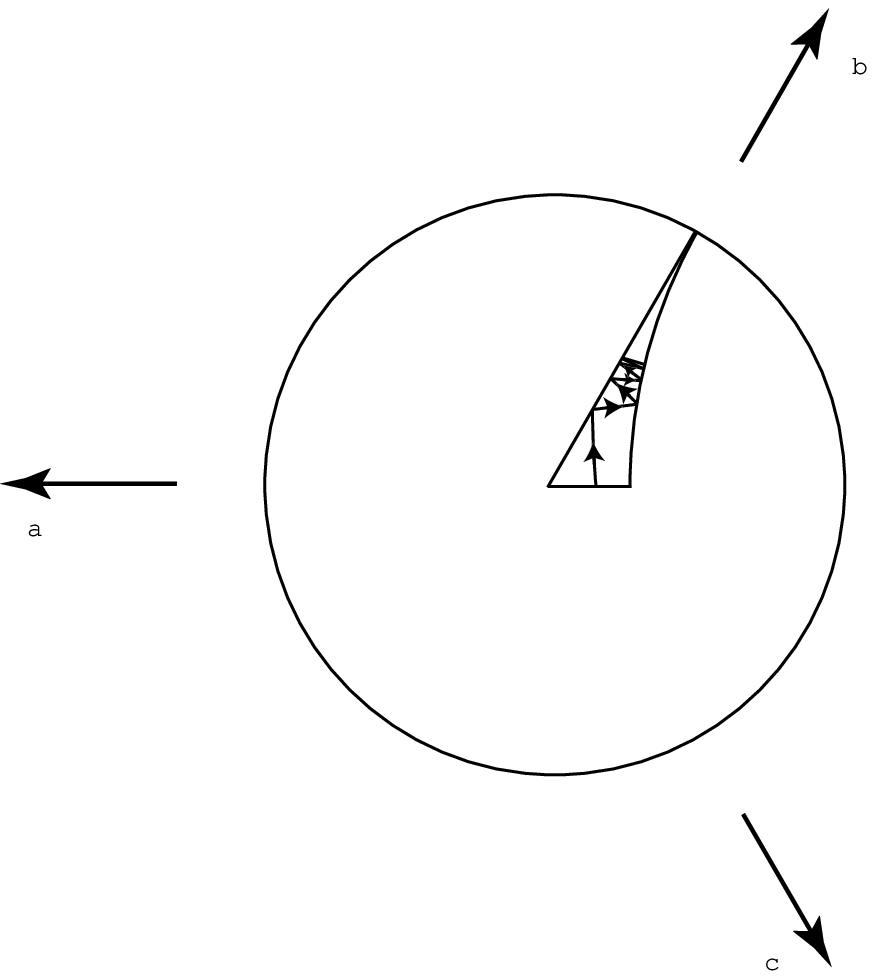}
\qquad\qquad & \qquad
\includegraphics[width=0.35\textwidth]{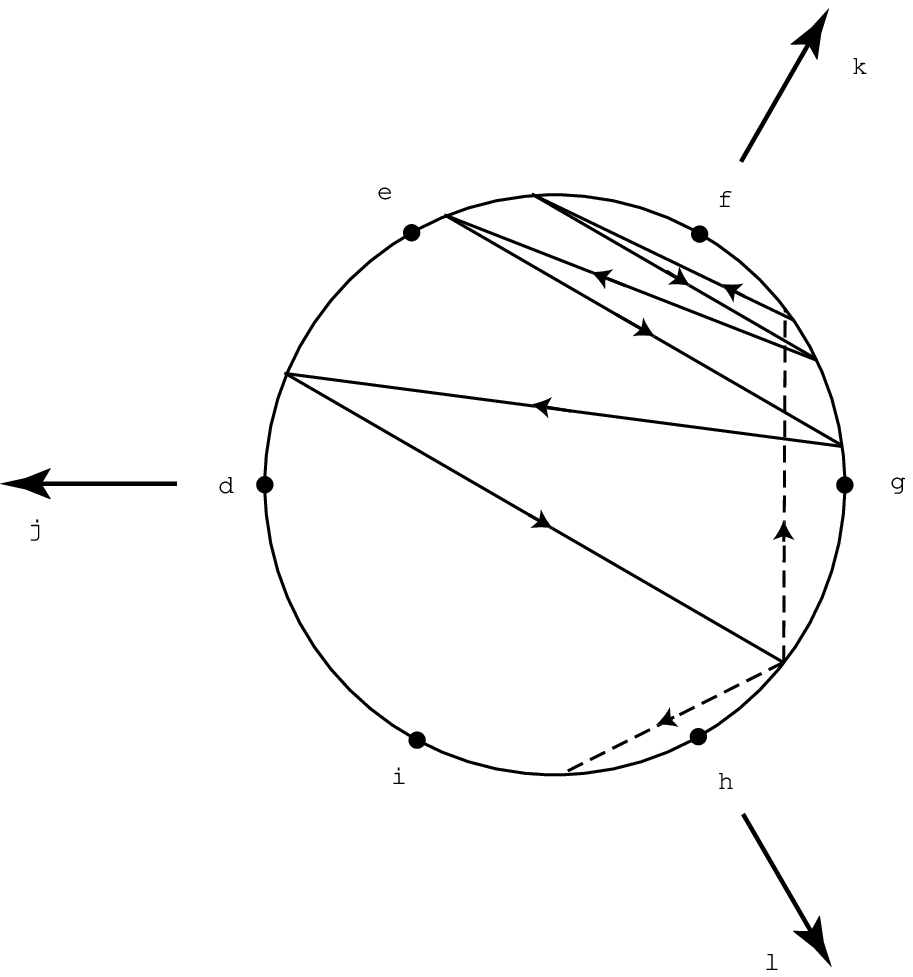}\\
(a) & (b)
\end{array}
$$
  \caption{Fig.~$(a)$ shows part of an orbit at a spatial point in terms of free
           Kasner (Fermi frame) motion and frame and curvature bounces. This represents
           a `configuration space' projection of the asymptotic dynamics. The disc here
           represents hyperbolic space. (The $\gamma^\alpha$ variables are `orthogonal'
           angular variables associated with a projection in
           $b^\alpha$-space, see~\cite{dametal03,heietal07}.) Fig.~$(b)$
           shows part of an orbit in terms of single frame and curvature transitions, i.e., it
           shows a part of an attractor sequence. Note that the solution does not quite return
           to any of the Kasner points it has `visited' before. This description
           represents a `momentum space' projection of the asymptotic dynamics. The circle here
           is the Kasner circle $\mathrm{K}^{\ocircle}$. The dashed lines correspond to the two
           possible single transitions that are possible at this stage; which one is realized depends
           on initial data. This corresponds to the situation that free motion in a given direction
           in Fig.~$(a)$ may lead either to hitting the wall associated with $R_1$ (the short wall)
           or with $N_1$ (the curved wall).}
     \label{dualfig}
\end{figure}

Thus the Hamiltonian `motion--bounce--motion--bounce' is replaced
with `bounce--motion--bounce--motion' in the dynamical systems
billiards. In the Hamiltonian approach the essential asymptotic
dynamical evolution is described by `configuration space' variables
(related to diagonalized spatial metric variables) while in the
dynamical systems approach it is described in terms of `momentum
space' variables ($\Sigma_\alpha$ are related to time derivatives of
the diagonalized spatial metric, and thus to the associated
momenta). The two approaches thus give \textit{complementary `dual'
representations} of the generic asymptotic dynamics; to compare the
two pictures, see Fig.~\ref{dualfig}. In addition the two approaches
mutually support each other and strengthen the credibility of the
generic picture; moreover, the combined use of both may very well be
of relevance for future developments, see~\cite{heietal07} for
further discussion.

\section{Concluding remarks}\label{concl}
In this paper we have presented the conformal Hubble-normalized
dynamical systems approach, and outlined the derivation of the
billiard attractor and established its `dual' correspondence with
the Hamiltonian cosmological billiard of Damour et
al.~\cite{dametal03}, but it is also of interest to make a
comparison with BKL: the generalized Kasner solution is obtained by
inserting the values of ${\bf S}$ at the Kasner circle
$\mathrm{K}^{\ocircle}$ into the equations for $E_\alpha{}^i$ and
$H$, and thus this `solution' is obtained as the lowest order
perturbation of $\mathrm{K}^{\ocircle}$ into the physical state
space ($E_\alpha{}^i \neq 0$); similarly the generalized Bianchi
type II solutions correspond to single curvature transitions
$\cT_{N_1}$---the ad hoc starting point of BKL is thereby derived.
Furthermore, as discussed, asymptotic (attractor) sequences connect
and yield precise meaning to BKL's `piecewise approximations,' and
thus the results of BKL's analysis are rigorously formulated as
special structures in the full extended dynamical systems picture.

Here we have been concerned with the vacuum case. However, the
methods employed in~\cite{heietal07} are equally applicable to many
other theories, see e.g.~\cite{dametal03,damnic05}, and to sources
as well. Regarding possible sources this naturally leads to a study
of the influence of matter on generic spacelike singularities, i.e.
structural stability of generic singularities, and this in turn
leads to a variety of issues, e.g., can the following be made more
substantial? Generic singularities seem to bring out the essential
properties of matter. There are indications that it is natural to
base a classification of the influence of matter on generic
singularity structure on (i) energy conditions and (ii) whether the
effective propagation speed is less than that of light or not. There
are also indications that suggest a subclassification based on how
`matter matters' in the light speed case, e.g. massless scalar
fields and electromagnetic fields influence the generic spacelike
singularity in different ways; does this motivate a
subclassification based on spin? If we assume the usual energy
conditions and consider the case of less than light speed
propagation, then `matter does not seem to matter' for generic
asymptotic dynamics towards a generic singularity---solutions are
said to be `vacuum dominated', or more precisely, the
Hubble-normalized energy-momentum tensor tends to zero so that the
geometry is asymptotically described by vacuum solutions. However,
this does not necessarily mean that e.g. the Hubble-normalized
rotation of a perfect fluid tends to zero, which suggests a
subclassification based on features that are (if any), or are not,
affected by matter. Another issue is how and if matter influences
the connection between generic spacelike singularities and weak null
singularities in asymptotically flat spacetimes (there are
indications that e.g. Vlasov matter behaves differently than perfect
fluids~\cite{heiugg06}); for special examples, see~\cite{limetal06}.

In this paper we have assumed that the dynamical evolution along
generic time lines is associated with asymptotic silence and
asymptotic local dynamics, which is supported by the results
in~\cite{heietal07}, but this does not mean that there could not
exist interesting phenomena associated with \textit{special} time
lines, indeed, we believe that a set of time lines of measure zero
exhibits spike formation and recurring `spike
transitions'~\cite{andetal05}, associated with non-local dynamics
and the formation of large spatial gradients---but asymptotic
silence still holds. In this context it is important to note that
even though spatial partial derivatives grow without bounds, the
Hubble-normalized spatial frame derivatives and the dimensionless
state space variables are still bounded, i.e., asymptotic
regularization still holds! Moreover, spike transitions seem to be
connected with asymptotically local dynamics and are governed by
variations of the Kasner map, hinting at further more deeply hidden
structures. Asymptotic spike formation is associated with the fact
that the unstable variables $N_1, R_1, R_3$ go through a zero at a
spatial surface, but the underlying mechanisms of spike formation
and possible spike annihilation are not completely
understood~\cite{lim06}: Where and why do spikes form? Do there
exist spikes that persist and lead to infinite sequences of spike
transitions (infinite recurring spike transitions)? Do spikes
annihilate spikes, and if so how: through `spike interference'? Is
spike formation more common than spike annihilation so that `spike
cascading' occurs? How many spikes form: do they form a dense set?
Do spikes---`gravitational defect-like' surfaces from the very early
universe---leave possible observational imprints in e.g. the CMB? It
clearly is of interest to answer these questions, moreover, a
clarification of some of these issues is of considerable interest in
the context of eventual generic singularity proofs.

Finally, one may ask why one should study generic singularities at
all in a classical GR context? Firstly, there exists a regime
between the Planck and GUT eras in the very early universe where GR
is expected to hold and where the approach towards the initial
singularity presumably is described by the dynamics towards a
generic singularity (recall that one of the points of inflation is
to `erase the effects of initial data,' and that before this erasure
a singularity is presumably generic according to this line of
reasoning). Secondly, black hole formation is connected with initial
data that reflects the complexities of the real universe; one would
hence also in this case expect generic spacelike singularities to
play a role before one enters the Planck regime. Thirdly, the
formation of generic singularities is associated with considerable
structure (for the exploitation of some of these structures in the
context of quantization of special models, see
e.g.~\cite{ashetal93,ashetal93b}), even in the case of spike
formation: Can this structure be used to asymptotically quantize
gravity where it needs to be quantized, namely in the ultra strong
gravitational field in the neighborhood of a generic spacelike
singularity?

\subsection*{Acknowledgements}
This research has been partially supported by the Swedish Research
Council. It is a pleasure to thank Lars Andersson, Henk van Elst,
Woei Chet Lim, Niklas R\"ohr, and John Wainwright for many helpful
and stimulating discussions about asymptotic silence and related
isssues; special thanks are due to Mark Heinzle for his crucial
contributions to~\cite{heietal07}, which forms the main foundation
for the present paper. Finally I would like to thank Bob Jantzen for
help with the manuscript, and especially for his support,
friendship, and fabulous cheesecakes.

\vfill

\end{document}